\documentclass[twocolumn,showpacs,preprintnumbers,amsmath,amssymb]{revtex4-1} 
\usepackage{graphicx}
\usepackage{ulem}
 \usepackage{amsmath}
\usepackage{amsbsy}
\usepackage{xcolor}%

\def\ni{\noindent}
\def\sig{\sigma}
\def\eps{\epsilon}

\def\br{\bf{r}}

\def\bff{\bf{f}}

\def\bs{\bf{s}}
\def\bt{\bf{t}}

\def\bF{\bf{F}}

\def\bR{\bf{R}}

\def\bC{\bf{C}}
\def\bP{\bf{P}}
\def\bQ{\bf{Q}}

\def\brho{\boldsymbol{\rho}}

\def\beps{\boldsymbol{\eps}}
\def\bsig{\boldsymbol{\sigma}}
\def\bdel{\bf{\nabla}}

\def\cA{{\cal A}}

\def\cF{{\cal F}}

\begin{document} 
 
 
\title{The unusual problem of upscaling isostaticity theory for granular matter}
\author{Raphael Blumenfeld}
 
\address{
Cavendish Laboratory, University of Cambridge, JJ Thomson Avenue, Cambridge CB3 0HE, UK} 

\date{\today} 

\begin{abstract} 
\ni Isostaticity theory (IT) provides stress field equations for marginally rigid packs of non-cohesive particles, interacting through hard-core forces. Its main advantage over strain-based theories is by closing the stress equations with stress-structure, rather than stress-strain, relations, which enables modelling the stress chains, often observed in experiments and simulations. 
The usefulness of IT has been argued to extend beyond its applicability at marginal rigidity. It has been shown to be derivable from first principles in two-dimensions, with the structure quantified by a fabric tensor that couples to the stress field. However, upscaling IT to the continuum is done currently empirically by postulating convenient closure equations.
The problem is that a volume average of the fabric tensor vanishes in the continuum limit, trivialising the closure equation. This poses an unusual upscaling problem, necessitating a new approach. Such an approach is developed here, resolving the problem for planar granular assemblies.
The new method is developed initially for idealised 'unfrustrated' packs by coarse-graining first to the two-grain scale, after which a conventional coarse-graining can be used. It is then extended to general realistic systems, by introducing an intermediate `de-frustration' procedure.
The applicability of the method is illustrated with a tractable example.

\end{abstract} 
\narrowtext

\maketitle 

\ni{\bf About Bob}:\\
Bob Behringer was one of the main players in the field of granular matter and contributed more than many to better understanding of these systems. When I first saw the beautiful images of force chains, produced by his experiments on photo-elastic particles \cite{BobPhotoElastic}, I remember thinking that this person had just tore a window in the wall of mystery surrounding the poorly understood granular systems. By devising a way to visualise force propagation paths in such media he made it possible to peer inside them and start to sort out the confusion surrounding them at the time. Indeed, his experiments gave rise to a new generation of predictive modelling. 

And then I met the man and realised that there was much more to Bob than his experiments. Bob impressed me immensely well beyond his science. The combination of his levelheadedness and his love of life was inspiring. Very notable was his default `cool, calm and collected' mode, with a pen between his teeth and a glint of humour in his eye. But that mode would last only until I described to him a potential new idea. The cool mode would disappear and he would react with the enthusiasm of a child. Indeed, his boundless curiosity made it easy to get him excited with good science. Beyond the science there were his abundant generosity and kindness. It was a pleasure to discuss with him anything, from fundamental science, through music and swimming to the meaning of life in general. Each such discussion felt like embarking on a voyage of discovery and almost always resulted in some more dots being connected and a better understanding emerging. Often, our discussions would meander into bantering, at which he was very good. It was very easy to forget how important his contribution had been to the development of the field and simply love him as a human being. 
Bob's death left a large gap in the field, as well as a hole in the hearts of many. The field is not the same without him. 

\bigskip
\ni{\bf Introduction}:\\
Granular materials (GMs) form an important and ubiquitous state of matter, playing a major role in our everyday life. They support and transmit stress significantly differently than conventional solids, often in the form of non-uniform stress chains {\cite{HuFi21}-\cite{Osetal06}. Having a fundamental theory to predict the stress that develops in such media under given boundary loads is extremely significant for a wide range of applications in fields beyond science and engineering. 
All continuum static stress theories are based on balance conditions of force and torque:
\begin{eqnarray}
\bdel\cdot\bsig &+ & \bF_{ext} = 0 \nonumber \\
\bsig & = & \bsig^T \ ,
\label{Balance} 
\end{eqnarray}
where $\bsig$ is the stress tensor, $\bsig^T$ its transpose, and $\bF$ includes all external and body forces on the medium. 
These equations must be supplemented by one constitutive relation in two dimensions (three in three dimensions). 
In elasticity theory, this is achieved by imposing compatibility conditions on the strain and relating the strain to the stress. This procedure results in elliptic equations whose solutions cannot be the observed stress chains. 
Alternative approaches, such as Critical State theory, popular in the engineering literature, rely on closures that relate the stress components when the medium is on the `yield surface', which is a surface in the space spanned by the stress components. This approach gives a hyperbolic set of equations, yielding stress chains, or slip lines, but it is phenomenological in nature and lacks predictive power \cite{Sc09}. 

At {marginal rigidity}, GMs are minimally connected, possessing the lowest possible mean number of force-carrying contacts per particle, at which the structure is at mechanical equilibrium. In this marginally rigid, or isostatic, state, the intergranular forces in the assembly can be determined from balance conditions alone. This makes the strain, and hence stress-strain relations, redundant as input at the grain level.
Since the continuum stress field is a coarse-grained representation of the spatial distribution of those forces then continuum strain-based constitutive properties are also redundant. For this reason, such theories, which includes elasticity, are inadequate for GMs. 

{Isostaticity stress theory has been developed to resolve the problem for ideally statically determinate, or isostatic, media \cite{Wietal96}-\cite{Geetal08}. Although most GMs are not precisely isostatic, on the verge of yielding they are sufficiently close to this state, containing isostatic regions, whose sizes depend on the proximity to marginal rigidity \cite{Bl04}. This makes isostaticity theory relevant to general GMs. Real systems can be generated close to marginal rigidity, e.g. by careful preparation or by generating low density shear bands \cite{BaBl03a,Bletal05}.  }

The irrelevance of strain for marginally rigid systems leaves structure as the only usable constitutive property for closing the stress equations 
{and isostaticity theory closes the equations by one stress-structure relation in 2D (three in 3D).}
The most general form of the relation in 2D is \cite{Wietal96}-\cite{EdGr99}
\begin{equation}
\sum_{ij} q_{ij}\sig_{ij} \equiv \bQ : \bsig = 0 \ ,
\label{Const}
\end{equation}
where $\bQ$ is a symmetric fabric tensor, quantifying the local structure in a specific way \cite{Fabric}. 
{Isostaticity theory is the name given to the stress field equations (\ref{Balance}) and (\ref{Const}). These equations are hyperbolic, resulting in solutions of stress chains extending along characteristic paths. While friction does not seem to appear explicitly in the equations, it affects the dynamics, that give rise to the static structure and, therefore, the statistics of the fabric tensor $Q$.}
Following empirical \cite{Wietal96, Wietal97} and mean field \cite{EdGr99} proposals for the form of $\bQ$, a first-principle theory derived it directly from the local grain-scale microstructure \cite{BaBl02}. Specifically, around a grain $g$ 
\begin{equation}
{\bQ}_g = \frac{1}{2} {\beps}^{-1}\cdot\left[{\bC}_g + \left({\bC}_g\right)^T\right]\cdot\beps \ ,
\label{Aii}
\end{equation}
where $\beps=\left(\begin{smallmatrix} 0 & 1 \\ -1 & 0\end{smallmatrix}\right)$ is a $\pi/2$-rotation matrix and  
\begin{equation}
{\bC}_g = \sum_c {\bC}_{cg} = \sum_c {\br}_{cg} \otimes {\bR}_{cg} \ ,
\label{Aiii}
\end{equation}
with the sum running over the loops $c$ that surround grain $g$. 
The quadrilateral, whose diagonals are the vectors ${\br}_{cg}$ and $\bR_{cg}$ (see Fig. \ref{fig:quadron}), is called 'quadron', and it is the structure's most basic volume element \cite{BlEd03}. For brevity, I index the quadrons in the following $cg$ by $q$ rather than $cg$.  
This structural tensor, which has a 3D equivalent \cite{BlEd06}, is useful for most GMs except when the grains are extremely non-convex \cite{BlEd07}. 
The quadrons surrounding grain $g$ are said to belong to it and their number equals its number of contacts with neighbouring grains.
This description allows us both to quantify the disordered local structure unambiguously and close the stress equations with a stress-structure coupling tensor ${\bQ}$.

It can be readily verified that Tr$\{{\bQ}_g\}=\sum_{q\in g}$Tr$\{{\bQ}_{q}\}=\sum_{q\in g} {\br}_{q}\cdot{\bR}_{q}$, where $q$ are the sum is over the quadrons belonging to grain $g$. Therefore, Tr$\{{\bQ}_g\}$ quantifies the deviation of the quadrons around grain $g$ from a kite form and, consequently, this is a measure of the net rotation of grain $g$ relative to a global mean, which must be zero \cite{BaBl02}. In other words, this quantity describes a local rotational fluctuation. 
It is constructive to consider
\begin{equation}
{\bP}_g = \beps {\bQ}_g {\beps}^{-1} = \sum_q \left( {\bs}_{q}{\bs}_{q} - {\bt}_{q}{\bt}_{q}\right)/2 \ ,
\label{PDef}
\end{equation}
where ${\bs}_{q}$ and ${\bt}_{q}$ are shown in Fig. \ref{fig:quadron}. 
The antisymmetric part of ${\bC}_{q}$ can be written as ${\cal{A}}\{{\bC}_{q}\}=A_{q}\beps$, with $A_{q}=\frac{1}{2}|{\br}_{q}\times{\bR}_{q}|$ being the area of the quadron $q$. The area associated with grain $g$ is $A_g=\sum_q A_{q} = {\cal{A}}\{{\bC}_g\}$ and the total area of the system is $A_{sys}=\sum_g A_g$.

\begin{figure}
\includegraphics[width=5cm]{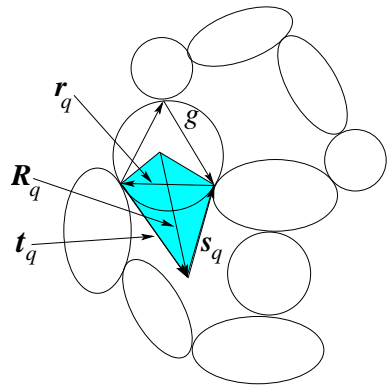}
\caption{The loop sides of the quadron $q$, whose diagonals are ${\br}_{q}$ and ${\bR}_{q}$ and whose area is $A_{q}$ (shaded), are the vectors ${\bs}_{q}$ and ${\bt}_{q}$. The symmetric tensor ${\bP}_g$ can be written in terms of these vectors, ${\bP}_g=\frac{1}{2}\sum_q\left({\bs}_{q}\otimes{\bs}_{q}-{\bt}_{q}\otimes{\bt}_{q}\right)$, with $q$ being all the quadrons belonging to grain $g$. Note that ${\bt}_{q}=-{\bs}_{q'}$ if $q'$ is adjacent to $q$. Therefore, the sum over ${\bP}_{q}$ inside a loop $c$ vanishes identically irrespective of the loop shape.}
\label{fig:quadron}
\end{figure}

\bigskip
\ni{\bf The problem}:\\
The first-principles derivation and, in particular, the ability to derive ${\bQ}_g$ from local structural characteristics \cite{BaBl02} were an important development, but a new problem emerged. The equations of a complete stress theory should have the same form when upscaled, which requires that ${\bQ}_g$ can be coarse-grained systematically to arbitrarily large length-scales. This is problematic because the volume-average of ${\bQ}_g$ over any region of space vanishes except for contributions from the region's boundary. This renders the closure relation (\ref{Const}) a useless trivial identity in the continuum limit. 
To see this, consider a region $\Gamma$, of boundary $\partial\Gamma$, area $A_{\Gamma}$, and $N_{\Gamma}\sim L^2$ grains. We have ${\bQ}_{\Gamma} = \langle {\bQ}_g\rangle_{\Gamma} = {\beps}\langle {\bP}_g\rangle_{\Gamma} {\beps}^{-1} = {\bP}_{\Gamma}$ and the volume-average of ${\bP}$ is

\begin{equation}
{\bP}_{\Gamma} = \frac{1}{2A_{\Gamma}} \sum_{q} \left( {\bs}_{q}\otimes{\bs}_{q} - {\bt}_{q}\otimes{\bt}_{q} \right) \ .
\label{Aiv}
\end{equation}
Let us partition the sum over neighbouring pairs in contact, $g$ and $g'$, surrounding loop $c$.  These terms cancel in pairs inside the loop because ${\bt}_{q\in c}=-{\bs}_{q'\in c}$.
It follows that the contribution of loops, fully enclosed within $\Gamma$, to the sum (\ref{Aiv}) vanishes - only that of the quadrons along the boundary $\partial\Gamma$ survives because those are not cancelled by neighbour quadrons. The vanishing of this sum is independent of the structural geometry, topology and grain shapes. It follows that ${\bP}_{\Gamma}\sim 1/L \to 0$ as $L$ increases, rendering (\ref{Const}) a trivial identity. 

{The method developed below consists of three steps. The first is by mapping the system into one with staggered order (SO), to be defined below. The second is coarse-graining the stress equations to the two-grain scale, with a fabric tensor that does not vanish identically to zero. The third step then proceeds as for any conventional coarse-graining method, by volume-averaging $Q$ over increasingly large volumes until the desired continuum limit is reached.}

\bigskip
\ni{The upscaling method}:\\
\ni{\bf I.} \underline{Packs with staggered order (SO)} \\
{The resolution of this problem is based on the observation that eq. (\ref{Const}), which was developed for frictional systems, can be rewritten in terms of only half the degrees of freedom in systems possessing a SO. } 
SO in granular packs means that grains can be labeled $+$ and $-$, such that any $+$ grain is in contact only with $-$ grains and vice versa. This is tantamount to the condition that each loop is surrounded by an even number of grains \cite{BaBl02}. 
{In such systems, the upscaling procedure is in the three steps detailed below: 
partitioning the assembly into pairs of $+$ and $-$ in contact; 
coarse-graining the constitutive equation to the two-grain scale by writing the stress equations for unit pairs in terms of half the degrees of freedom; 
upscaling the constitutive equation by conventional volume averaging.}

The general force moment on grain $g$ is $\cF_g = \sum_{h}{\brho}_{gh}\times{\bff}_{gh}$, where $h$ are the neighbours of $g$, ${\bff}_{gh}$ the force that grain $h$ exerts on $g$, and ${\brho}_{gh}$ are the position vectors of the contacts between $g$ and $h$. The local stress on grain $g$ is then ${\bsig}_g = {\cF}_g / A_g$. 
The mean and differential stresses of a +/- pair are, respectively, ${\bsig}_m = (\cF_+ + \cF_-)/A_{pair}$
and ${\bsig}_d = (\cF_+ - \cF_-)/A_{\pm}$, where $A_{\pm} = A_+ + A_-$ is the area associated with the pair. 
Averaged over a large region, $\Gamma$, $\langle{\bsig}_m\rangle_{\Gamma}$ converges to the traditional continuum stress field, with the local stress fluctuations 
$\langle{\bsig}_d\rangle_{\Gamma} \sim 1/L \to 0$, with $L$ the the system linear size. 
We combine the constitutive equations (\ref{Aii}) for the pair, rearrange:
\begin{eqnarray}
({\bQ}_+ + {\bQ}_-):{\bsig}_m + ({\bQ}_+ - {\bQ}_-):{\bsig}_d & = & 0 \nonumber \\
({\bQ}_+ - {\bQ}_-):{\bsig}_m + ({\bQ}_+ + {\bQ}_-):{\bsig}_d & = & 0 \ ,
\label{Av}
\end{eqnarray}
and average over $\Gamma$. The first equation becomes a trivial identity. The correlation between the two small quantities, $\langle{\bsig}_d\left({\bQ}_+ + {\bQ}_-\right)\rangle_{\Gamma}$ also decays with size, which leaves $\langle({\bQ}_+ - {\bQ}_-):\sig_m\rangle = 0$, in which the subscript $\Gamma$ was dropped for brevity. 
The averages of $\bQ$ and $\bsig$ decouple as $\Gamma$ increases: 
\begin{equation}  
\langle({\bQ}_+ - {\bQ}_-):\sig_m\rangle = \langle{\bQ}_+ - {\bQ}_-\rangle :\langle\sig_m\rangle \ .
\label{Avi}
\end{equation}
Since $\langle {\bQ}_+ + {\bQ}_-\rangle = 0$ then $\langle {\bQ}_+\rangle =- \langle{\bQ}_-\rangle$ and we can substitute this into (\ref{Avi}) to obtain the upscaled relation 
\begin{equation}
\langle{\bQ}_{+} \rangle :\langle{\bsig}_m\rangle = 0 \ .
\label{eq:Diii}
\end{equation}
This constitutive relation is a consistent coarse-grained version of the original one.
Moreover, it has the significant advantage that the average of the constitutive quantity ${\bQ}_+$ does not vanish identically on coarse-graining like the original equation. Thus, this resolves the coarse-graining problem in SO systems. 

{The method presented next aims to coarse-grain the fabric tensor in packings of frictional grains. Packings of ideally frictionless grains have been shown in \cite{BaBl02} to be mappable naturally to systems of frictional grains that possess SO. This is done by first assuming that the grains are frictional and then introducing infinitesimally small ball bearings at the contact points. It has been shown that the resultant system remains isostatic. Thus, for frictionless systems, eq. (\ref{eq:Diii}) can be coarse-grained straightforwardly by volume averaging and the following procedure is not required.}

\bigskip
\ni{\bf II.} \underline{Extension to general GMs}\\
In general GMs, there are almost always loops surrounded by odd numbers of grains (OL), which `frustrates' the SO, as at least two same-sign grains must be in contact.

\begin{figure}
\includegraphics[width=6cm]{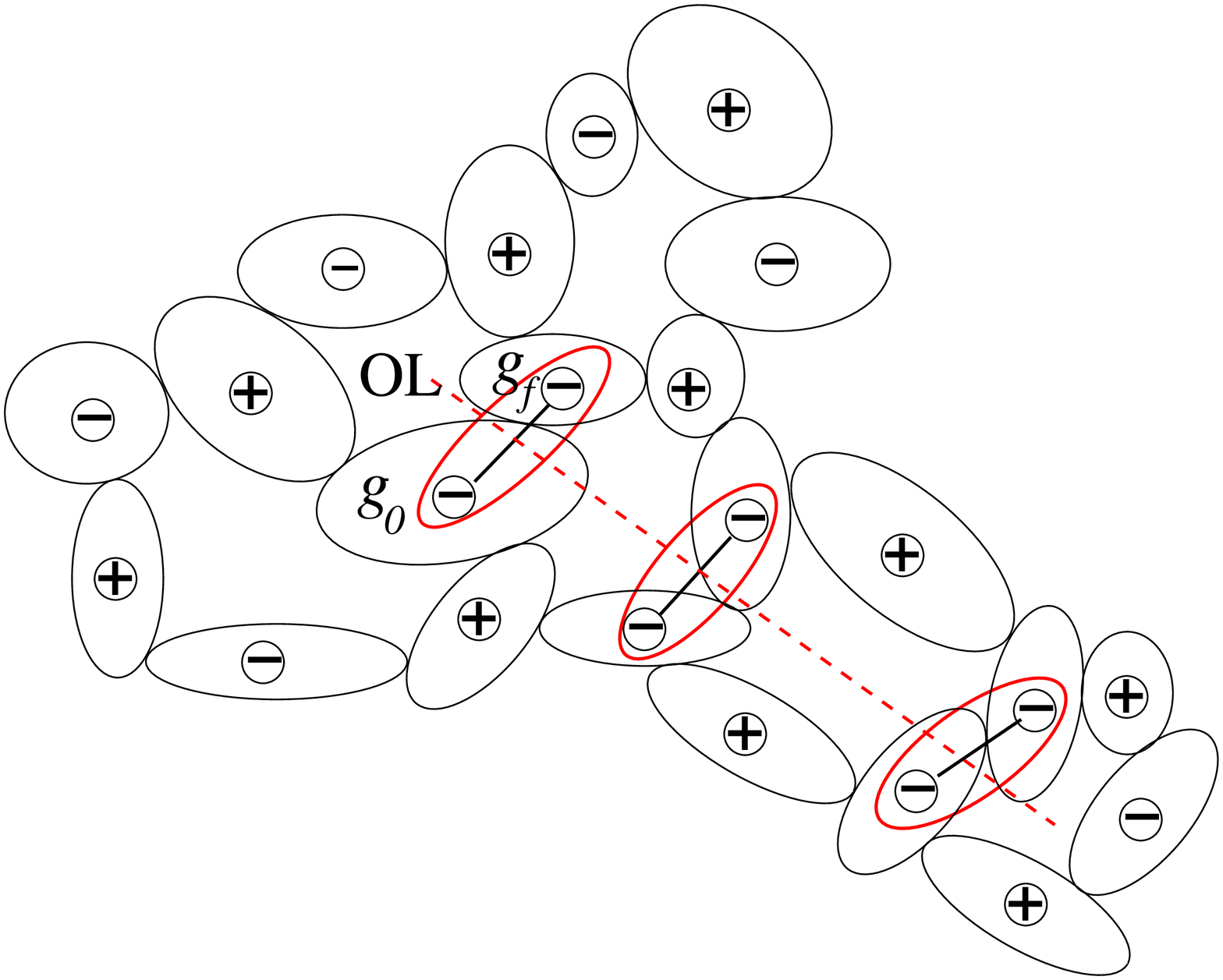} 
\caption{The frustration around an OL surrounded by even-edged loops. Labelling the grains around it alternately by + and -, leaves two neighbouring 'frustrated' grains with the same sign ($g_0$ and $g_f$). Starting from a neighbour of $g_0$ and labelling the next shell (thick lines), leaves another frustrated pair lying adjacent to the first one. Continuing this procedure shell by shell around the original OL, results in a line of such frustrated grain pairs emanating from it. This line is capped when it is incident on another OL.}
\label{fig:FrustLine} 
\end{figure} 

The extension of the above method to general systems is by renormalising the structure to lift the frustration. 
{It is convenient to focus first on systems without 3-edge loops - the extension to include these will be described below.}
Consider a single OL, within a region of only even-edged loops, such as sketched in Fig. \ref{fig:FrustLine}. Label the grains around it as + and - in the clockwise direction, alternatingly. The last grain, $g_f$, is in contact with the first one, $g_0$, and they are both $-$. Starting from a neighbour of $g_0$, which does not belong to the OL,  label similarly the first shell of loops surrounding the OL. With this shell's loops being even-edged, it must contain exactly one frustrated pair of grains and these, which is adjacent to the pair $g_0$-$g_f$. Repeating this process shell by shell outwards, a continuous line of same-sign pairs emanates from the single OL, as sketched in Fig. \ref{fig:FrustLine}. This line extends to the system boundary. 
However, if the line is `incident' on another OL, it ends, as sketched in Fig. \ref{fig:FrustLine}. 

This observation is the key to renormalising the structure for lifting the frustration. Isolating same-sign grain pairs into lines and regarding each such pair as a rigid super-grain, we recover a SO structure!  The procedure is the following. Firstly, identify all the OLs in the network. Next, partition the OLs into nearest pairs, minimising the overall number of contacts between them. Thirdly, identify a `frustration' line between each OL pair, using the above method and avoiding crossing of these lines. Finally, declare each frustrated pair of grains a super-grain and compute its fabric tensor, $Q$ as the sum of the fabric tensors of its constituents. The result is a system of rigid objects, some the original grains and some the super-grains, possessing a SO. Finally, use the above coarse-graining procedure to upscale the constitutive relation to regions of required size, namely, containing sufficient numbers of loops each. 

\begin{figure}
\includegraphics[width=9cm]{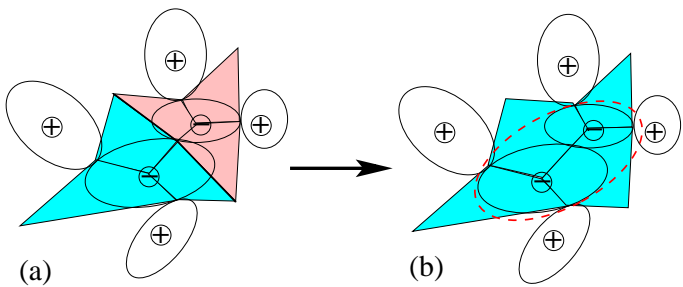} 
\caption{The frustrated pair in (a) is joined into a super-grain, denoted by a dashed red line, in (b). The area of the super-grain is the sum of the areas of its constituents, which is also the coefficient of the antisymmetric part of the renormalised geometric tensor $\bC_{pair}$. 
Since the renormalised fabric tensor ${\bP}$ is the sum over the tensors of the frustrated grains, their common vectors $\bs$ and $\bt$ between the joined pair (thick black lines in (a)) are also eliminated in (b), disposing of the effect of the contact point in the structural description. The stress field is unaffected by this because these vectors cancelled out anyway in the original structure. Therefore, the joining procedure is self-consistent.}
\label{fig:FabricRenorm} 
\end{figure} 
Although this renormalisation changes the local topology, it does not alter the forces on the original grains. Turning two grains into a super-grain eliminates the contact between them and consequently some quadrons, which are the descriptors and basic elements of the local structure. 
Since the structure tensor of the super-grain is the sum of those of its constituents, ${\bC}_{\pm}={\bC}_+ + {\bC}_-$, then the area associated with the super-grain is $A_{\pm}\beps = \cA\{{\bC}_{\pm}\} = \left(\cA\{{\bC}_+\} + \cA\{{\bC}_-\}\right) = \left(A_+ + A_-\right)\beps$, as expected.
Another feature of the procedure concerns the physics of the stress field. Considering eq. (\ref{Aiii}), eq. (\ref{Aiv}) and Fig. \ref{fig:FabricRenorm}, note that the vectors $\bs$ and $\bt$, which originally extended between the eliminated contact point and the centres of the loops flanking it, cancel out and they do not play any role in the renormalised tensor ${\bQ}_{\pm}$ that results from the pair of the eliminated grains. Since the stress-structure coupling can be mediated only by grain contacts, the removal of the effect of the eliminated contact point by the local renormalisation is significant - it underpins the self-consistency of the procedure with its local structural and geometrical interpretation. 

{3-edge loops are special because joining any two of their grains annihilates the loop altogether. Such loops are very sparse in isostatic frictional systems, whose mean number of loop edges is 6 (the result of Euler's theorem). These loops are pre-processed before the upscaling to the two-grain scale, by merging the loop's three grains into an initial super-grain. In the rare occasion that clusters of such 3-edge loops occur the entire cluster is merged into a super-grain. The de-frustrastion procedure is then applied to the resulting structure, which is devoid of 3-edge loops.}

{Once the 3-edge loops have been eliminated, the procedure described in the paper works for any concentration of odd loops and, in particular, when the entire system is made of them. Such systems are highly unlikely to occur but, if they do, the upscaling procedure would result in the two grains, shared by every pair of neighbouring loops, being joined  into a super-grain. This loses an edge for each cell, making them all even-edged, and reduces the number of loops to a half. }

\bigskip
\ni{\bf Example}:\\
To illustrate the method, consider first the granular pack shown in Fig. \ref{fig:GranCat}. The underlying structure is ordered on a honeycomb-like lattice, possessing SO. 
A local defect has been introduced by expanding one loop into an octagon, generating two pentagonal neighbours. The latter are OLs, introducing two sources of local frustration. The structure tensor $\bC$ is purely antisymmetric in the regular hexagonal regions and $\bQ_g$ vanishes. The frustration defects, highlighted by ellipses in the figure, introduce a local finite fluctuation in the fabric tensor and we wish to calculate the renormalised effect on the constitutive relation after renormalising to SO. Carrying out the above joining procedure on the frustrated grains, and summing over the resulting $\bQ$ tensors of the positive grains, we obtain the renormalised contribution:
\begin{equation}
{\bQ}_{defect}=-\left(\begin{smallmatrix} {\frac{3}{10}}&{\frac{4\sqrt{3}}{5}} \\ {\frac{4\sqrt{3}}{5}}&{\frac{3}{2}} \end{smallmatrix}\right) \ . 
\label{QDefect}
\end{equation}
Substituting into (\ref{eq:Diii}), the local constitutive relation around the defect is then
\begin{equation}
\sig_{xx} + 5\sig_{yy} + \frac{16}{\sqrt{3}}\sig_{xy} = 0 \ .
\label{Div}
\end{equation}
\begin{figure} 
\includegraphics[width=6cm]{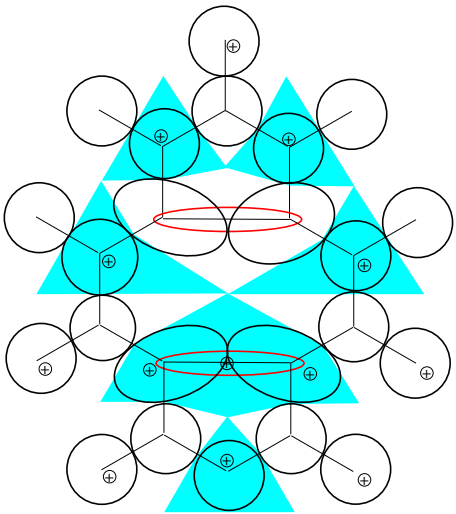} 
\caption{A granular Kagome structure, based on a honeycomb lattice, is ordered except for a defect consisting of an octagonal loop flanked by two pentagonal 
loops, which introduce two frustrated grain pairs. The coarse-graining is over all the positive grains, but only those near the defect with irregular areas (shown shaded) have a finite fabric tensor.}
\label{fig:GranCat} 
\end{figure} 
Generally, adding a uniformly random distribution of defects at area density $\rho$ to a defect-free isostatic system, of an initial fabric tensor ${\bQ}_0$, the coarse-grained fabric tensor of the effective medium is ${\bQ}_{em}=(1-\rho){\bQ}_0+\rho{\bQ}_{defect}$. Using this constitutive relation together with (\ref{Div}) provides the stress equations at any desired length scale.

It is useful to illustrate the issue with a system that is partly uniform and partly with defects. Suppose the uniform medium is the above honeycomb, deformed slightly to perturb the lattice's symmetry, which trivialises the diagonal of ${\bQ}_0$. Placing particles in contact on the vertices makes a Kagome structure. The deformation can be chosen to yield a range of fabric tensors of the form ${\bQ}_0=Ka^2\left(\begin{smallmatrix} {1-\xi}&{-(1+\xi)} \\ {-(1+\xi)}&{1-\xi} \end{smallmatrix}\right)$, with K a constant and $\xi<1$ \cite{HaBl11}. For simplicity, I choose a deformation that yields $\xi=1/4$. In the uniform medium, this gives rise to characteristics oriented at gradients $\lambda^{0}_1=1/3$ and $\lambda^{0}_2=3$ ($A$ and $B$ in Fig. \ref{EMTSolution}). Let this medium fill the entire half-plane $-\infty\leq y\leq\infty$ and $0\leq x\leq \infty$. Now fill the the region $x_0\leq x\leq\infty$ with the aforementioned defects at density $\rho = 10\%$, the fabric tensor in this region is
\begin{equation}
{\bQ}_{em}(x>x_0)=C^{te} \left(\begin{smallmatrix} {1}&{9+2.56\sqrt{3}} \\ {9+2.56\sqrt{3}}&{4.44} \end{smallmatrix}\right) \ . 
\label{Qeff}
\end{equation}
The resulting characteristics in this region are at gradients $\lambda^{em}_1\approx0.17$ and $\lambda^{em}_2\approx26.70$.

Applying at the origin the stress $\sigma_0=\left(\begin{smallmatrix} {s_{xx}}&{0} \\ {0}&{0} \end{smallmatrix}\right)$ gives rise to stresses along the left two characteristic paths, $A$ and $B$: $\sigma_{A}=\frac{-3s_{xx}}{8}\left(\begin{smallmatrix} {1}&{3} \\ {3}&{9} \end{smallmatrix}\right)$ and $\sigma_{B}=\frac{27s_{xx}}{8}\left(\begin{smallmatrix} {1}&{1/3} \\ {1/3}&{1/9} \end{smallmatrix}\right)$. 
These chains reach the boundary of the defect-filled region at points $(x_0,x_0/3)$ and $(x_0,3x_0)$, whereupon they act as boundary loads for the medium in $x>x_0$. From each of these load sources emanates a pair of stress chains at gradients $\lambda^{em}_1$ and $\lambda^{em}_2$. The stresses along these paths is calculated using the method outlined in \cite{Geetal08}. Fig. \ref{EMTSolution} illustrates $\sigma_{xx}/s_{xx}$ along the chains $C,D,E,F$: $-1.06275, -8.88825, 0.56475, 88.99425$, respectively.

\begin{figure} 
\includegraphics[width=6cm]{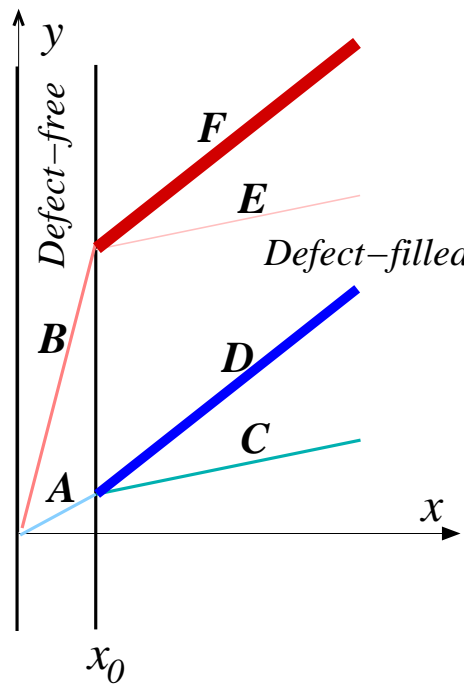} 
\caption{The stress chain paths (gradients are not to scale for better illustration) in the defect-free ($x<x_0$) and defect-filled ($x\geq x_0$) regions. The  blue (red) represent negative (positive) values of $\sigma_{xx}$. The darker the shade and the wider the line the higher the stress magnitude. }
\label{EMTSolution} 
\end{figure}

\bigskip
\ni{\bf Conclusion}:\\
In summary, the problem of upscaling the stress equations of isostatic GMs is unique in that the volume average of the constitutive fabric tensor, which couples to the stress to close the field equations, vanishes in the large-scale limit. This has led to imposing phenomenological or empirical fabric tensors. Particular examples of such closures are the yield conditions, such as Mohr-Coulomb, in {plasticity-based theories}. This problem is resolved here from first-principles by developing a specialised upscaling method. The method is based on the observation that the constitutive relation can be written in terms of only half the degrees of freedom in ideally unfrustrated granular packs, whose volume average over $Q$ need not vanish identically in the continuum limit. {Using this observation, the closure equation is first upscaled to the two-grain scale in such systems and from then on coarse-grained to the continuum conventionally by volume averaging.}

{Since most granular structures are frustrated, a `de-frustration' method has been developed to transform any planar granular structure into an unfrustrated one. The method is based on joining frustrated grain pairs to lift the local frustration.} 
The procedure leads to a renormalisation of the local fabric tensor and an example of such a calculation for a simple system including two defects in an otherwise honeycomb-like structure has been illustrated. 
{It should be noted that the `de-frustrated' structure need not be, and in most cases is not, marginally rigid. However, this does not pose a difficulty because the original physical system is marginally rigid and therefore isostaticity theory applies regardless of the mathematical manipulation of the structure.}

{The unusual aspects of this upscaling procedure are a direct consequence of the vanishing of a straightforward volume averaging of the constitutive quantity $Q$, a feature that is not common in any other coarse-graining procedure. This difficulty necessitates an upscaling in several stages: (i) de-frustrating the system into one with SO; (ii) upscaling to the two-grain scale by writing the closure equation for pairs of grains and using half the degrees of freedom; (iii) conventional volume averaging of the renormalised $Q$ over increasing lengthscales.}

It would be interesting to test the method on systems in which both the structure and the forces can be visualised, such as the many sample systems produced in the lab of Bob Behringer \cite{BobPhotoElastic}. Moreover, the problem is still outstanding for three-dimensional systems, which have not been discussed here. Work on extension of the method to such systems is ongoing.

\bigskip
\ni{\bf Compliance with ethical standards}:\\
I declare that I have no conflict of interest.


\begin{thebibliography}{99}
\bibitem{Fabric} {Note that several other descriptors in the literature are named `fabric tensor', but those are mostly useful for strain-based theories, with the conditions for being useful discussed in \cite{LiDa15}. Since isostaticity theory is not strain-based, these conditions do not apply, while $Q$ is useful by construction, being based on the contact network alone. }
\end{thebibliography}
\end{document}